\documentclass[10pt,letterpaper]{article}
\usepackage[top=0.85in,left=2.75in,footskip=0.75in]{geometry}

\usepackage{hyperref}
\hypersetup{colorlinks,linkcolor={red},citecolor={blue},urlcolor={blue}} 
\usepackage{amsmath,amssymb}

\usepackage{changepage}

\usepackage[english, vietnamese]{babel}

\usepackage{textcomp,marvosym}

\usepackage{cite}

\usepackage{nameref,hyperref}

\usepackage{microtype}
\DisableLigatures[f]{encoding = *, family = * }

\usepackage[table]{xcolor}

\usepackage{array}

\usepackage{tabularx}
\usepackage{multirow}
\usepackage{subfigure}

\newcolumntype{+}{!{\vrule width 2pt}}

\newlength\savedwidth

\raggedright
\usepackage[aboveskip=1pt,labelfont=bf,labelsep=period,justification=raggedright,singlelinecheck=off]{caption}

\makeatletter
\renewcommand{\@biblabel}[1]{\quad#1.}
\makeatother

\usepackage{lastpage,fancyhdr,graphicx}
\usepackage{epstopdf}
\fancyheadoffset[L]{2.25in}
\fancyfootoffset[L]{2.25in}
\begin{document}
\begin{otherlanguage}{english}

\vspace*{0.2in}

\begin{flushleft}

{\Large
\textbf\newline{Learning to diagnose common thorax diseases on chest radiographs from radiology reports in Vietnamese}
}
\\
Thao T.B. Nguyen\textsuperscript{1,\Yinyang}, Tam M. Vo \textsuperscript{1,\Yinyang}, Thang V. Nguyen\textsuperscript{1}, Hieu H. Pham\textsuperscript{1,2,3,*}, Ha Q. Nguyen\textsuperscript{1,2}
\\
\bigskip
\textbf{1} Smart Health Center, VinBigData JSC, Hanoi, Vietnam
\\
\textbf{2} College of Engineering and Computer Science, VinUniversity, Hanoi, Vietnam
\\
\textbf{3} VinUni-Illinois Smart Health Center, Hanoi, Vietnam
\\
\bigskip
%
%
\Yinyang These authors contributed equally to this work.





\textsuperscript{*}Corresponding author (Hieu H. Pham). Email: \textcolor{blue}{hieu.ph@vinuni.edu.vn}
\end{flushleft}
\section*{Abstract}

Deep learning, in recent times, has made remarkable strides when it comes to impressive performance for many tasks, including medical image processing. 
One of the contributing factors to these advancements is the emergence of large medical image datasets. However, it is exceedingly expensive and time-consuming to construct a large and trustworthy medical dataset; hence, there has been multiple research leveraging medical reports to automatically extract labels for data.
The majority of this labor, however, is performed in English.
In this work, we propose a data collecting and annotation pipeline that extracts information from Vietnamese radiology reports to provide accurate labels for chest X-ray (CXR) images.
This can benefit Vietnamese radiologists and clinicians by annotating data that closely match their endemic diagnosis categories which may vary from country to country. To assess the efficacy of the proposed labeling technique, we built a CXR dataset containing 9,752 studies and evaluated our pipeline using a subset of this dataset. With an F1-score of at least 0.9923, the evaluation demonstrates that our labeling tool performs precisely and consistently across all classes. After building the dataset, we train deep learning models that leverage knowledge transferred from large public CXR datasets. We employ a variety of loss functions to overcome the curse of imbalanced multi-label datasets and conduct experiments with various model architectures to select the one that delivers the best performance. Our best model (CheXpert-pretrained EfficientNet-B2) yields an F1-score of 0.6989 (95\% CI 0.6740, 0.7240), AUC of 0.7912, sensitivity of 0.7064 and specificity of 0.8760 for the abnormal diagnosis in general. Finally, we demonstrate that our coarse classification (based on five specific locations of abnormalities) yields comparable results to fine classification (twelve pathologies) on the benchmark CheXpert dataset for general anomaly detection while delivering better performance in terms of the average performance of all classes.

\section*{Introduction}

Radiography has always been one of the most ubiquitous diagnostic imaging modalities so far, while chest X-ray (CXR) is the most commonly performed diagnostic X-ray examination~\cite{bib1}. CXRs has an important role in clinical practice, effectively assisting radiologists to detect pathologies related to the airways, pulmonary parenchyma, vessels, mediastinum, heart, pleura and chest wall~\cite{bib2}. In recent years, great advances in GPU computing and research in the fields of machine learning have led to the trend of automating CXR image diagnostics~\cite{bib47,bib48,bib39,bib49,bib40,bib50,bib43,bib51} and many other X-ray modalities~\cite{bib37,bib52,bib38,bib41}. In addition, the availability of large-scale public datasets~\cite{bib3,bib4,bib42,bib5,bib6,bib7} has sparked interest in study and application, with some of them already being used and integrated into the Computer-Aided Diagnosis (CAD) system to reduce the rate of CXR misdiagnosis.

Several datasets, including CheXpert~\cite{bib3}, MIMIC-CXR~\cite{bib4}, PadChest~\cite{bib5}, Chest-xray8, Chest-xray14~\cite{bib6} and VinDr-CXR \cite{bib7,bib46}, VinDr-PCXR~\cite{bib44,bib45}, had a significant impact on increasing labeling methods and model quality. Building a reliable CXR dataset for a specific project, on the other hand, remains a difficult and challenging task because medical data is difficult to obtain due to numerous restrictions on patient information confidentiality, and label quality is heavily influenced by the doctors' experience and subjective opinion~\cite{bib1}. This is costly and time-consuming but essential, especially for a task that tackles specific challenges, such as focusing on a certain set of patients or illnesses. In such a way that adopting the aforesaid large-scale datasets is sometimes ineffective, possibly because the image quality, labeling, or data characteristics are no longer appropriate. Additionally, CXR images and medical reports corresponding to each examination are also stored in hospital storage systems such as Picture Archiving and Communication System (PACS) and Hospital Information System (HIS) during the radiology process. This is a tremendous available resource to build large-scale CXR datasets in which the annotation can be automatically interpolated from the free text report without any involvement of radiologists. Therefore, pipelines or methods to create datasets from available resources are always valuable.

Some previous works also developed methods to relabel public large datasets or constructed a new one. Wang et al.~\cite{bib6} proposed a method for extracting a hospital-scale CXR dataset from the PACS via an unified weakly-supervised multi-label image classification and disease localization formulation by applying natural language processing (NLP) techniques. NegBio~\cite{bib8}, a rule-based algorithm that utilizes universal dependencies and subgraph matching, known as providing regular expression infrastructure for negation and uncertain detection in radiology reports. Filice et al.~\cite{bib9} investigated the benefit of utilizing AI models to create annotations for review before adjudication in order to speed up the annotation process while sacrificing specificity. Johnson et al.~\cite{bib4} extracted and classified mentions from the associated reports using two NLP tools, CheXpert and NegBio, before aggregating them to arrive at the final label. To construct structured labels for the images, Irvin et al.~\cite{bib3} created an automated rule-based labeler to extract observations and capture uncertainties contained in free-text radiology reports.  Padchest~\cite{bib5} labeled the majority of the dataset using a recurrent neural network with an attention mechanism. This dataset contains excerpts from Spanish radiology reports, however the labels have been mapped to biological vocabulary unique identifier codes, making the resource useful regardless of the language. RadGraph~\cite{bib10} introduced a new dataset of clinical entities and relations annotated in full-text radiology reports taken from CheXpert and MIMIC.  This research made use of a novel information extraction schema that extracts clinically relevant information associated with a radiologist’s interpretation of a medical image.

More advanced NLP approaches, such as Bidirectional Encoder Representations from Transformers (BERT)~\cite{bib36}, are used in some studies. Chexpert++~\cite{bib11}, a BERT-based, high fidelity approximation labeler applied to CheXpert, is significantly faster, fully differentiable, and probabilistic in outputs. VisualCheXbert~\cite{bib12} utilized a biomedically-pretrained BERT model to map directly from a radiology report to the image labels, with a supervisory signal determined by a computer vision model trained to detect medical conditions from chest X-ray images. CheXbert~\cite{bib13} is a BERT-based approach to medical image report labeling that exploits both the scale of available rule-based systems and the quality of expert annotations. 

Dictionary-based heuristics are another popular way for creating structured labels from free-text data. For instance, MedLEE~\cite{bib14} utilizes a pre-defined lexicon to convert radiology reports into a structured format. Mayo Clinic’s Text Analysis and Knowledge Extraction System (cTAKES)~\cite{bib15} tool combines dictionary and machine learning methods, and uses the Unified Medical Language System \footnote{\url{https://www.nlm.nih.gov/research/umls/index.html}} (UMLS) for dictionary inquiries. Dictionary-based NLP systems have a key flaw is that they do not always establish high performance when handling in-house raw clinical texts, especially those with misspellings, abbreviations, and non-standard terminology. On top of that, the mentioned systems only cover English language and cannot handle non-English clinical texts. Languages other than English, including Vietnamese, do not have sufficient clinical materials to build a medical lexicon. In nations where English is not the official language, this has been a huge obstacle in building clinical NLP systems. In current work, our data pipeline can be applied for the available data in PACS and HIS, which can assist minimize data labeling costs, time, and effort while reducing radiologists' involvement in the workflow. We propose a set of matching rules to convert a typical radiology report to the normal/abnormal status of classes. 

Other than the above-mentioned differences in labeling methods, our label selection is also different from previous studies. So far, most of the studies were developed for classifying common thoracic pathologies or localizing multiple classes of lesions. For instance, most deep learning models were developed on the MIMIC-CXR~\cite{bib16} and CheXpert~\cite{bib17,bib18,bib19} datasets for classifying 14 common thoracic pathologies on CXRs in recent years. The earlier dataset ChestX-ray14~\cite{bib20}, an expansion of ChestX-ray8~\cite{bib21}, including the same set of 14 findings has been used to develop deep learning models~\cite{bib22,bib23}. 
Nevertheless, these approaches are far different from how Vietnamese radiologists work. In clinical practice, a CXR radiology report always includes four descriptions that correlate to four fixed anatomical regions of the thorax: chest wall, pleura, pulmonary parenchyma and cardiac. 
Therefore, it is not practical for Vietnamese radiologists to utilize a CAD system that provides suggestions for the presence of 14 diseases. Typically, when examining a CXR image, radiologists analyze that image by region; consequently, it is more convenient for the system to indicate the abnormality of each area, eliminating the need to match the lesion type with the region being viewed.
To address the realistic demand of Vietnamese radiologists, we developed a system to classify CXRs into 5 classes depending on the position of pathologies: chest wall, pleura, parenchyma, cardiac abnormality and the existence of abnormalities in the CXRs, if any. When tested on the benchmark CheXpert dataset, we found that this coarse classification produces results comparable to the detailed classifier of 14 findings in terms of abnormal class and gives better results in terms of macro average F1 score of all classes. 

Our work was developed on the dataset collected at Phu Tho General Hospital - a Vietnamese provincial hospital. To develop trainable images with corresponding labels, DICOM files in PACS are matched with radiology reports retrieved from HIS. By extracting data from radiology reports, generating normal/abnormal status of 5 classes and treating it as the ground-truth reference, we can conclude that there were positive results when classifying CXRs according to 5 groups of pathologies, which are modeled after the radiologist’s description in their medical report. Unlike the automatic data labeling methods mentioned above, our proposed method is simple yet accurate by filtering the descriptions alluding to no findings first, then searching for phrases implying abnormalites in each position. Therefore, the labeling process is strictly controlled through stages, making it easy to detect errors and correct them. In addition, adding a manual step to the labeling process helps us deal with misspellings, which was neglected by the previous method. In this step, we also find infrequent phrases, adding them to our list of phrases indicating abnormality to make it more complete. Furthermore, a report always includes descriptions corresponding to four fixed anatomical regions of the thorax, thus by generating set of labels matching these regions, we can minimize the chance that a label is uncertain.

\section*{Material and method}
\subsection*{Dataset building pipeline}

Our proposed pipeline consists of five steps: (1) data collection, (2) PA-view filtering, (3) XML parser, (4) data matching and (5) data annotation. Fig~\ref{fig1} illustrates the above five steps in detail. Firstly, DICOM files stored in PACS will be acquired and filtered to retain only posterior-anterior (PA) view CXRs by the PA classifier  application programming interface (API). Meanwhile, radiology reports stored in HIS as XML files will be parsed to attain some specific information. Afterward, DICOM files and radiology reports belonging to the same patient will be matched to generate pairs of DICOM-XML files of the same examination. Once a DICOM file has been determined to match with an XML files, that DICOM file will be converted to JPG format and the XML file will be the subject of a labeling tool to generate a set of corresponding labels. At the end of the procedure, we can obtain a trainable dataset which includes JPG images and their corresponding labels.

\begin{figure}[th!]
\centering
\includegraphics[width=1.\linewidth, scale=1.]{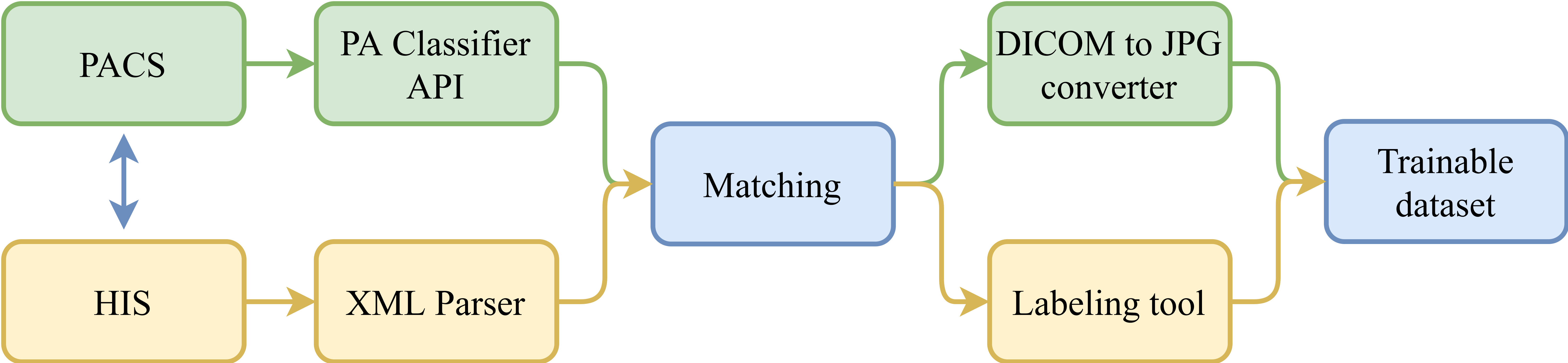}
\caption{{\bf Overview diagram of the process of collecting and building medical image dataset.} 
The process consists of five steps: data collection from PACS and HIS, PA-view filtering, XML parser, data matching and data annotation.}
\label{fig1}
\end{figure}

\subsubsection*{Data collection}

We retrospectively collected chest radiography studies from Phu Tho General Hospital, which were performed within five months from November 2020 to March 2021, along with their associated radiology reports. The ethical clearance of these studies was approved by the Institutional Review Board (IRB) of Phu Tho General Hospital. With this approval, the IRB allows us to access their data and analyze raw chest X-ray scans using our VinDr's platform, which will be used for data filtering. The need for obtaining informed patient consent was waived because this retrospective study did not impact clinical care or workflow at the hospitals, and all patient-identifiable information in the data has been removed.

We decided to select four types of pathologies because of their prevalence in the medical reports and clinical practice. An example of a typical description extracted from a radiology report is shown in Fig~\ref{fig2}. The description is divided into four main categories: lungs, cardiac, pleura and chest wall by most Vietnamese radiologists. 
From the four groups of pathology, we create an annotation set consisting of five classes, with the first four classes corresponding to these four groups and the other indicating the presence of abnormalities on CXRs, if any.

\begin{figure}[ht!]
\centering
\subfigure[Vietnamese radiology description]
    {\includegraphics[width=0.49\linewidth]{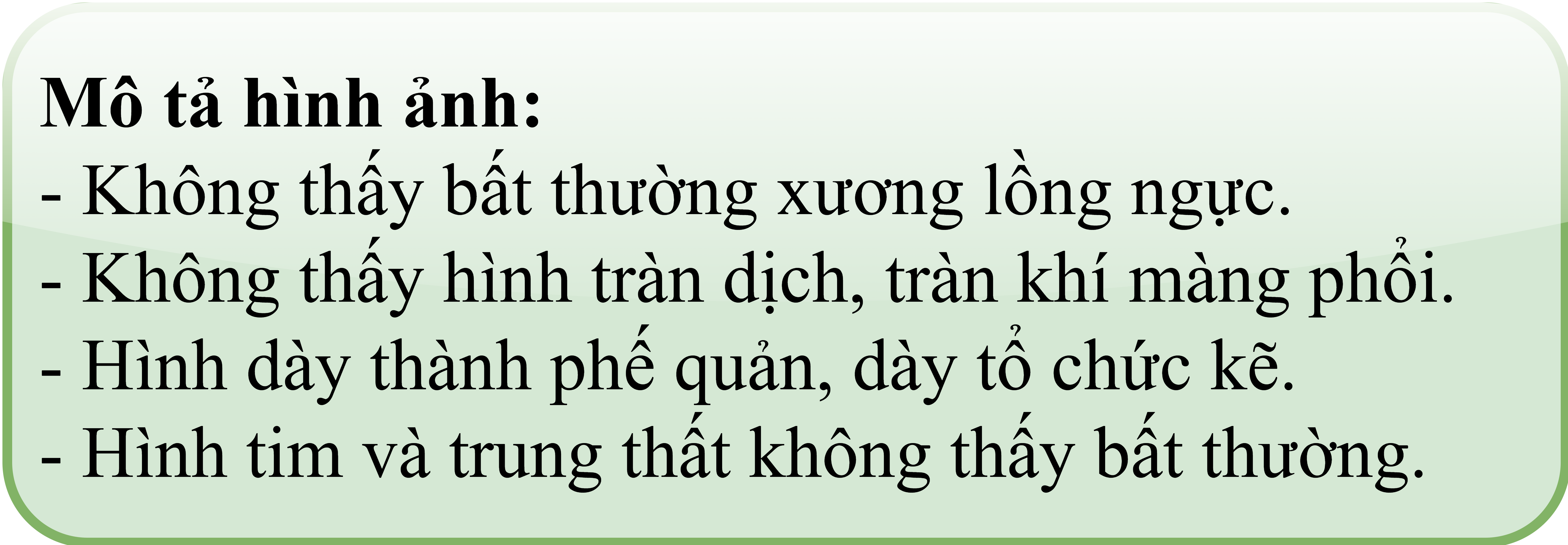}} \hfill
\subfigure[Translation of Vietnamese radiology description]
    {\includegraphics[width=0.49\linewidth]{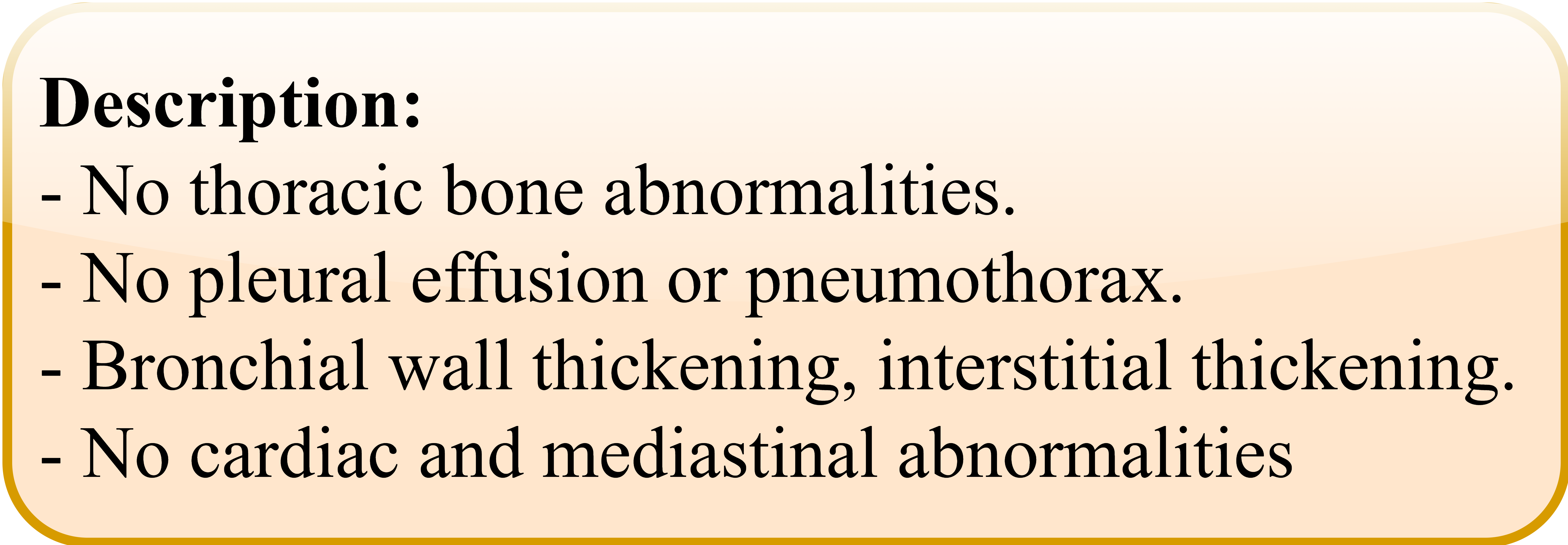}}
\caption{{\bf The description in a typical radiology report in Vietnam.}
The description is divided into four main categories: chest wall, pleura, lungs (parenchyma) and cardiac.}
\label{fig2}
\end{figure}

\subsubsection*{PA-view filtering}

The collected data was mostly of Posterior-Anterior (PA)-view CXR, but also included a large number of outliers such as images of body parts other than chest, low-quality images or images with different views than PA-view. To guarantee that only CXRs of PA-view will be retained, we ran an API that is powered by VinDr's platform~\footnote{\url{https://vindr.ai/vindr-lab}}. The API takes a DICOM file as an input and returns the probability that the image saved in that file is a PA-view CXR. The DICOM file will proceed to the next stage of data pre-processing if this probability exceeds 0.5 - a normalized threshold; else, the file will be marked as ignored.

\subsubsection*{XML parser}

We use the same procedure for the XML parsing and data matching process as in our previous study~\cite{bib31}, shown in Fig~\ref{fig3}. The figure illustrates the procedure of extracting radiology reports from HIS. Each assessment and treatment session was saved in the Extensible Markup Language (XML) file format by HIS. A session includes all information of the patient between check-in and check-out time. The XML parser can read the header of a session that includes SESSION\_ID, PATIENT\_ID, CHECK\_IN\_TIME, and CHECK\_OUT\_TIME. These attributes are shared among all radiology reports belonging to the same session and will be used to link to the corresponding DICOM file. All reports are also interpreted using the XML parser to obtain the SERVICE\_ID, REPORT\_TIME, and DESCRIPTION properties. Only reports with a SERVICE ID matching the values expressly assigned by the Vietnamese Ministry of Health for chest radiography were preserved to exclude extraneous reports.\\

\begin{figure}[ht!]
\centering
\includegraphics[width=0.9\linewidth, scale=0.8]{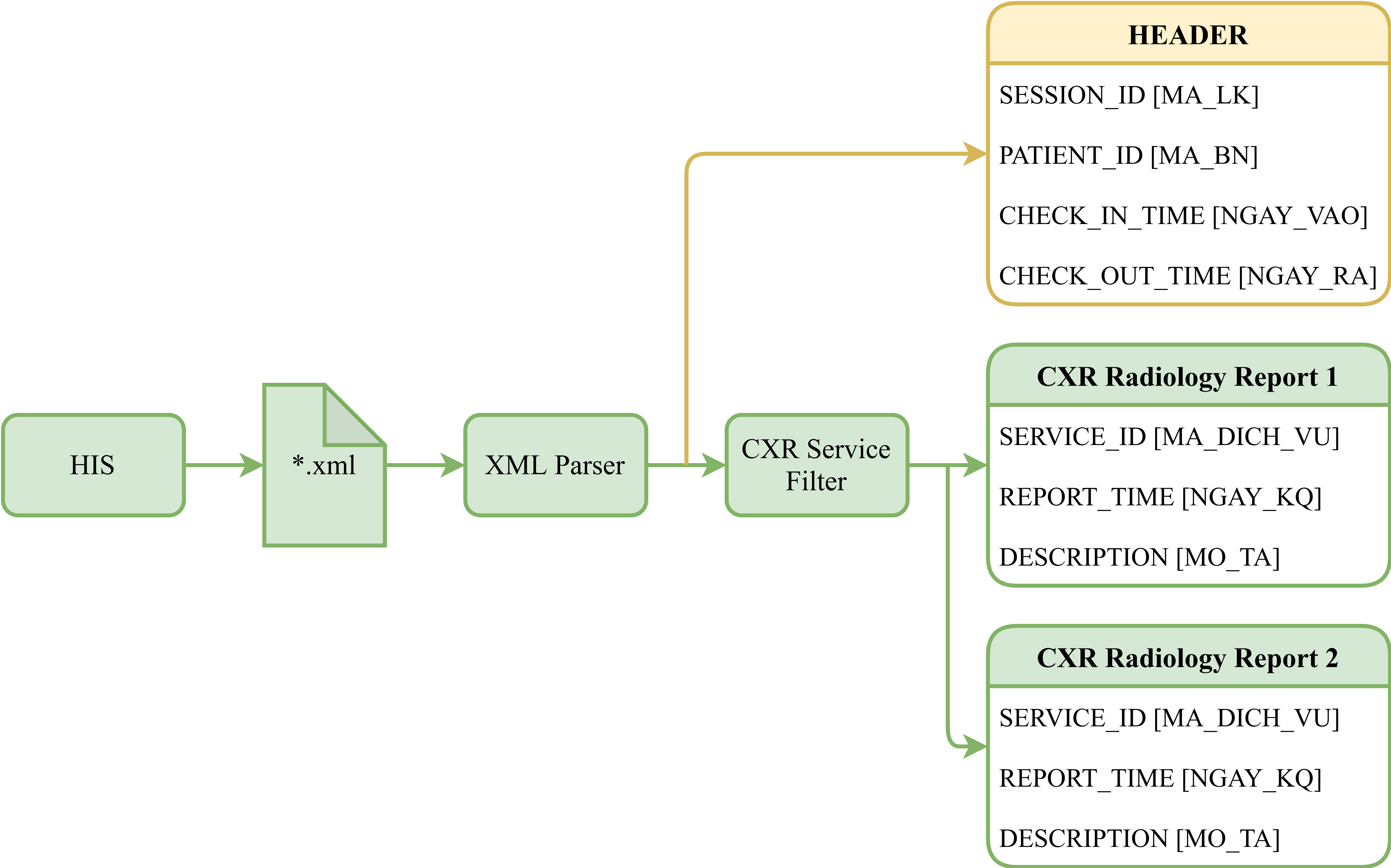}
\caption{{\bf Radiology reports extraction process for CXR examinations collected from HIS~\cite{bib31}.} 
The original Vietnamese counterparts are put inside square brackets.}
\label{fig3}
\end{figure}

\subsubsection*{Data matching}

To match the DICOM file with the corresponding XML file, we have simulated the algorithm in~\cite{bib31}, which is depicted in Fig~\ref{fig4}. Since the HIS and PACS are linked by PATIENT\_ID, this key is used by the matching algorithm to determine whether the DICOM file and radiography report belong to the same patient. Moreover, REPORT\_TIME must be within 24 hours of STUDY\_TIME, which is a regulated protocol of the hospital. Finally, STUDY\_TIME has to be between CHECK\_IN\_TIME and CHECK\_OUT\_TIME. If all of the conditions are fulfilled, the DICOM file and the radiology report are matched.

\begin{figure}[ht!]
\centering
\includegraphics[width=0.75\linewidth, scale=0.2]{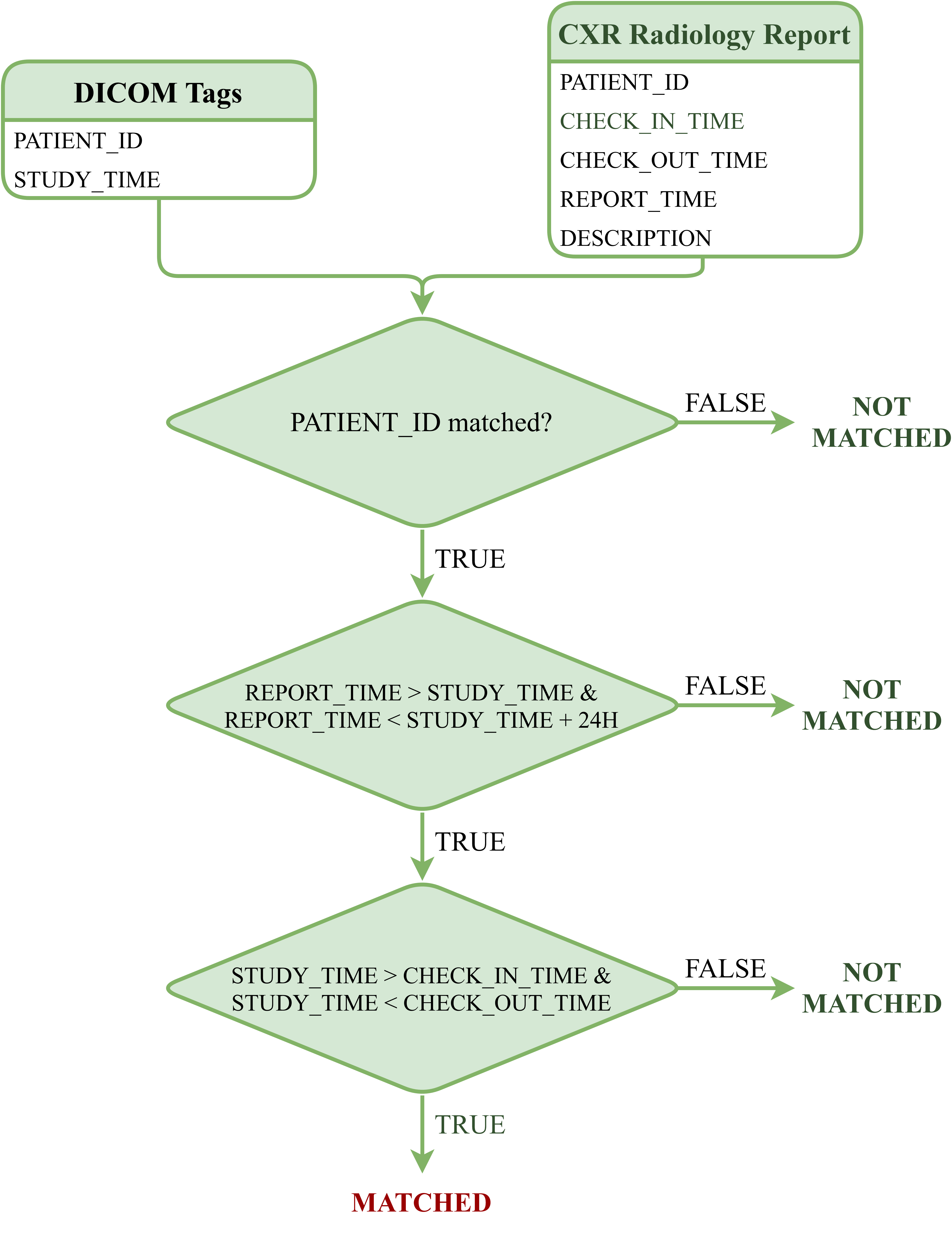}
\caption{{\bf Algorithm for matching a DICOM file obtained from PACS with a radiology report collected from HIS.}}
\label{fig4}
\end{figure}

One problem we encountered here is that one DICOM file matched multiple reports and vice versa, because their STUDY\_TIME attributes were separated by a period of less than 24 hours. In such a short period of time, the examination results are often the same, the reason for taking additional radiographs may be due to the poor quality of the first image. Therefore, the description from the reports is usually the same, and this DICOM file is assigned to one of the matched reports. In several cases where the descriptions in the reports are different, the DICOM file will be given to a radiologist to review and match the correct report.

\subsubsection*{Data annotation}

After extracting descriptions that match the DICOM files, we developed a simple labeling algorithm that takes the radiologists' description as input and returns a list of five binary elements, corresponding to the presence or absence of abnormalities belonging to 5 classes. Fig~\ref{fig5} illustrates the major steps of data annotation, which is implemented in semi-automated manner, including (1) pattern filtering, (2) keyword detection (3) abnormality interpolating and (4) manually labeling.

\begin{figure}[ht!]
\centering
\includegraphics[width=\linewidth, scale=0.8]{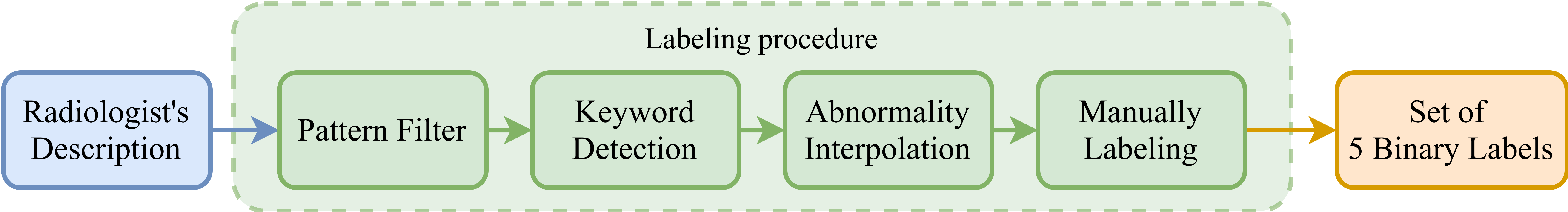}
\caption{{\bf Semi-automated data annotation pipeline. }The system consists of 4 steps, the first 3 steps are automatic and the last one is carried out manually.}
\label{fig5}
\end{figure}

{\bf Pattern Filtering} The dataset we obtained from Phu Tho General Hospital is unbalanced, with the majority of the images exhibiting no pathology. We have obtained 1,568 different templates from all the descriptions. Filtering descriptions that are elements of the predetermined set of templates (specifically 11 templates imply no findings) would help us save a significant amount of time when it comes to data labeling. A CXR is considered normal if one of 11 templates exactly appears in the DESCRIPTION of the corresponding radiology report.

{\bf Keyword detection} After pattern filtering, most of the instances without pathologies are retained. In this step, we have to handle most of the abnormality descriptions and some remaining normality ones. Keyword detection is divided into four sub-stages, which could be performed simultaneously, to detect keywords indicating abnormalities in the chest wall, pleura, parenchyma, and mediastinum. To find keywords for each class, e.g. chest wall, we break down the radiologist’s description into 4 categories (categories are separated by ``-" (dash) in the radiology descriptions). From the sentences in the chest wall category, we gather keywords indicating abnormalities, such as ``fracture", ``osteoporosis", ``bone fusion surgery" to create the fixed set of keywords. Descriptions containing keywords in the chest wall set will be annotated as 1 for the corresponding class, similarly for the pleura, parenchyma, and cardio classes. 
Some common keywords setting for the four classes are listed in Table~ \ref{table1}.

\begin{otherlanguage*}{vietnamese}
\begin{table}[!ht]
    \centering
    \scriptsize{
    \caption{{\bf Examples of Vietnamese keywords indicate abnormalities in chest wall, pleura, parenchyma, cardiac classes and abnormality  out of these four group.} English translations are enclosed in square brackets.}
    \label{table1}
    
    \begin{tabularx}{0.8\linewidth}{|c|>{\raggedright\arraybackslash}X|}
        \hline
        {\bf Class name} & {\bf Keywords} \\ 
        \hline
        
        \multirow{3}{*}{Chest wall (bone)} & Gãy xương [Bone fracture] \\ & Thưa xương [Osteoporosis] \\ & Tiêu xương [Bone resorption] \\
        \hline
        
        \multirow{3}{*}{Pleura} & Dày màng phổi trái{/}phải [Left{/}right pleural thickening] \\ & Mờ góc sườn hoành màng phổi trái{/}phải [Left{/}right costophrenic angle blunting] \\ & Tù góc sườn hoành trái{/}phải [Loss of the left{/}right costophrenic angle]\\
        \hline
        
        \multirow{3}{*}{Parenchyma} & Dày thành phế quản [ Bronchial wall thickening] \\ & Dày tổ chức kẽ [Interstitial pulmonary thickening] \\ & Dải mờ giữa phổi trái{/}phải [Opacity between left{/}right lung \\
        \hline
        
        \multirow{3}{*}{Cardio} & Quai động mạch chủ (đmc) vồng [Ascending aortic arch] \\ & Hình tim trái{/}phải to [Enlarged {/}right cardiomegaly] \\ & Giãn cung thất trái{/} phải [Left{/}right ventricular arch dilatation] \\
        \hline
        
        {Other abnormality} & Liềm hơi dưới vòm hoành trái{/}phải [Sickle of air below the left{/}right diaphragm] \\ 
        \hline
    \end{tabularx}}
\end{table}

\end{otherlanguage*}

{\bf Abnormality interpolating} The first four classes have been annotated at the keyword detection stage, here the abnormality class labeling is implemented by inferring from those others. Abnormality value will be set to 1 (positive) if any of the other classes are noted as anomalies or has any other anomaly even though it does not belong to the four groups above.

{\bf Manual Labeling} Descriptions that neither belong to the 11 normality templates nor contain any of the keywords in the four fixed sets have a high probability of being misspelled or describing rare pathologies or including pathologies that cannot be assigned to one of the four main regions. To handle such cases, we inspected them to correct spelling mistakes manually, then forwarded confusing descriptions to a radiologist of Phu Tho General Hospital for annotating. These cases account for less than 0.5\% of the total descriptions, thus labeling the remain is not a time-consuming task, that minimizes the doctor's involvement in data labeling.

Over five months, we obtained the total number of 12.367 XML files and 12,376 DICOM files coresponding to 11,088 studies. 10,847 DICOM files were PA chest radiographs, and 10,002 of them matched with information extracted from XML files. Table \ref{table2} details the number of positive and negative samples of the five classes in the collected dataset. For model development, we split the dataset into training and validation sets with the ratio of 7/3 and one constraint is that the distribution of each class in training and validation sets is approximated to the distribution of the original dataset.

\begin{table}[!ht]
\centering
\scriptsize{\caption{
{\bf Number of instances which contain five labeled observations in training, validation and the whole dataset.}}
\label{table2}

\begin{tabularx}{\linewidth}{|c|>{\centering\arraybackslash}X|>{\centering\arraybackslash}X|}
    \hline {\bf Position of pathology} & {\bf Positive} & {\bf Negative} \\ 
    \hline {\bf Chest wall} 
        & {
        \begin{tabularx}{\linewidth}{l>{\raggedright\arraybackslash}X}
            Training & 166 \\
            Validation & 71 \\
            Total & 237 (2.37\%)
        \end{tabularx}}
        & {
        \begin{tabularx}{\linewidth}{l>{\raggedright\arraybackslash}X}
            Training & 6835 \\
            Validation & 2930 \\
            Total & 9765 (97.63\%)
        \end{tabularx}} \\
    \hline {\bf Pleura} 
        & {
        \begin{tabularx}{\linewidth}{l>{\raggedright\arraybackslash}X}
            Training & 155 \\
            Validation & 67 \\
            Total & 222 (2.22\%)
        \end{tabularx}}
        & {
        \begin{tabularx}{\linewidth}{l>{\raggedright\arraybackslash}X}
            Training & 166 \\
            Validation & 71 \\
            Total & 9780 (97.78\%)
        \end{tabularx}} \\
    \hline {\bf Parenchyma} 
        & {
        \begin{tabularx}{\linewidth}{l>{\raggedright\arraybackslash}X}
            Training & 1520 \\
            Validation & 652 \\
            Total & 2172 (21.72\%)
        \end{tabularx}}
        & {
        \begin{tabularx}{\linewidth}{l>{\raggedright\arraybackslash}X}
            Training & 6846 \\
            Validation & 2934 \\
            Total & 7830 (78.28\%)
        \end{tabularx}} \\
    \hline {\bf Cardio} 
        & {
        \begin{tabularx}{\linewidth}{l>{\raggedright\arraybackslash}X}
            Training & 548 \\
            Validation & 235 \\
            Total & 783 (7.83\%)
        \end{tabularx}}
        & {
        \begin{tabularx}{\linewidth}{l>{\raggedright\arraybackslash}X}
            Training & 6453 \\
            Validation & 2766 \\
            Total & 9219 (92.17\%)
        \end{tabularx}} \\
    \hline {\bf Abnormal} 
        & {
        \begin{tabularx}{\linewidth}{l>{\raggedright\arraybackslash}X}
            Training & 1976 \\
            Validation & 848 \\
            Total & 2824 (28.23\%)
        \end{tabularx}}
        & {
        \begin{tabularx}{\linewidth}{l>{\raggedright\arraybackslash}X}
            Training & 5025 \\
            Validation & 2153 \\
            Total & 7178 (71.77\%)
        \end{tabularx}} \\ 
        \hline
\end{tabularx}}
\end{table}

\subsection*{Quality control}

To ensure the quality of the dataset is guaranteed, we randomly take 5\% of the data to inspect if there are any inappropriate images or labels that do not match the corresponding report. If any incorrectness is found, we will find out and correct it, then the 5\% selection process is repeated until no more errors are detected. The inspection was carried out by a medical student majoring in radiology and was double checked by a radiologist of Phu Tho General Hospital.

\subsection*{Labeler results}
We evaluate the effectiveness of the proposed labeling procedure by manually labeling the samples and considering the result as the ground truth. F1-score will be used as the main metric to evaluate the quality of our labeling tool.
\subsubsection*{Evaluation Set}
The reported evaluation set consists of 3001 radiology reports from 3001 instances - that totally overlap with the reports in the validation set.
We manually annotated these radiology reports without access to additional patient information. We labeled whether there is any abnormality in chest wall, pleura, pulmonary parenchyma and cardio following a list of labeling conventions that was agreed upon ourselves. After we independently labeled each of the 3001 reports, disagreements were resolved by consensus discussion or radiologist's consultation. The resultant annotation serves as ground truth on the report in evaluation set.
\subsubsection*{Evaluation results}
After having the results as the radiologists' annotation, combined with the set of labels generated by our method, the evaluation results of each class are listed in Table~\ref{table3}, with the metrics of precision, recall and F1 score. Overall, our labeling pipeline delivers the high values of F1 score in all classes, with the lowest figures of 0.9926 and 0.9985 - being recorded in pleura and parenchyma classes, respectively. In chest wall, cardio and abnormal classes, our tool delivers the highest performance, without any mislabeled samples.

\begin{table}[!ht]
\centering
\scriptsize{\caption{
{\bf Evaluation results of proposed labeling tool.} Evaluation was performed on 3001 samples of the validation set.}
\label{table3}
\begin{tabularx}{\linewidth}{|c|*{7}{>{\centering\arraybackslash}X|}}

\hline
{\bf Class} & {\bf TP} & {\bf FP} & {\bf TN} & {\bf FN} & {\bf Precision} & {\bf Recall} & {\bf F1 score}\\ 
\hline

Chest wall & 71 & 0 & 2930 & 0 & 1 & 1 & 1 \\
\hline

Pleura & 67 & 1 & 2933 & 0 & 0.9853 & 1 & 0.9926 \\
\hline

Parenchyma & 652 & 1 & 2347 & 1 & 0.9985 & 0.9985 & 0.9985 \\
\hline

Cardio & 235 & 0 & 2766 & 0 & 1 & 1 & 1 \\
\hline

Abnormal & 848 & 0 & 2153 & 0 & 1 & 1 & 1 \\
\hline

\end{tabularx}}
\end{table}

\section*{Experiment and results}

\subsection*{Model development}

Chest X-ray interpretation with deep learning methods usually relies on pre-trained models developed for ImageNet. Nevertheless, it was proved that architectures achieving remarkable accuracy on ImageNet are unlikely to give the same performance when experienced on the CheXpert dataset and the choice of model family deliver better improvement than image resizing within a family for medical imaging tasks~\cite{bib24}. We decided to choose the model family that has been proved to be highly efficient for CXR interpretation - ResNet50~\cite{bib32}, DenseNet121~\cite{bib33}, Inception-V3~\cite{bib34} and EfficientNet-B2~\cite{bib35}. We also leverage large public CXR datasets such as CheXpert to develop pre-trained models and compare the use of some benchmark chest X-ray datasets for transfer learning to ImageNet pre-trained models. Furthermore, the unbalance between classes has a negative impact on our dataset; for example, the chest wall class has a positive/negative ratio of 0.003. To address this problem, along with the conventional Binary Cross Entropy Loss (BCE), we used and assessed other loss functions established for multi-label imbalanced datasets, such as Asymmetric Loss (ASL)~\cite{bib25} and Distribution-balanced Loss (DBL)~\cite{bib26}.

For each model architecture, we use the Adam optimizer (beta1 = 0.9, beta2 = 0.999 and learning rate = 1e-3), cooperating with Cosine annealing learning rate with gradual warm-up scheduler, a batch size of 16, three different loss functions: cross-entropy, distribution-balanced and asymmetric loss, image sizes of 768 and 1024.

Training was conducted on a Nvidia GTX 1080 with CUDA 10.1 and Intel Xeon CPU ES-2609. For one run of a specific model, we train for 160 epochs and evaluate each model every 413 gradient steps. Finally, checkpoint with the highest F1-score will be considered the best model for each training procedure. 

We also used the nonparametric bootstrap~\cite{bib27} to estimate 95\% confidence intervals for each statistic. There are 3,000 replicates are drawn from the validation set, and the statistic is calculated for each replicate. This procedure generates a distribution for each statistic, by reporting the 2.5 and 97.5 percentiles, the confidence intervals are obtained and significance is assessed at the p = 0.05 level.

\subsection*{Experimental result}

In this work, chest X-ray classification models were trained on the training set detailed in Table 2. The models are distinguished from each other based on four attributes: (1) model architecture, (2) pre-trained dataset, (3) loss function and (4) image size, while sharing the common training procedure. First, we compare the effect of using pre-trained datasets and the impact of some loss functions on the multi-label problem. We choose ImageNet and CheXpert to transfer their knowledge to our target data. BCE - a common loss function, ASL and DBL - the two loss functions for multi-label issue were used in our experiment. The reported metrics are macro average (Av.) F1-score, AUC, sensitivity and specificity of the five classes. We only use ResNet50 architecture to compare these aspects with the same setup hyper parameters.

As we can see in Table~\ref{table4}, model using ASL and CheXpert dataset as pre-trained-initial parameters give the best result. All the metrics are higher than that of the others, especially when using ASL. This loss function always gives big value but is very effective because it heavily ``penalizes" misclassified positive samples and hardly penalizes easy negative one. CheXpert is also useful in spite of containing similar patterns to our target data. We decide to use pre-trained model by CheXpert and ASL for later experiments.

\begin{table}[!ht]
\centering
\scriptsize{\caption{{\bf Experimental results with different pre-train datasets and loss functions.} Model pre-trained on CheXpert dataset and using Asymmetric loss function yields the best performance.}
\label{table4}
\begin{tabularx}{\linewidth}{|c|*{5}{>{\centering\arraybackslash}X|}}
	\hline
	{\bf Pretrained dataset} & {\bf Class} & {\bf F1 score} & {\bf AUC} & {\bf Sensitivity} & {\bf Specificity} \\
	{\bf + Loss function} &  &  &  &  & \\
	\hline
	\multirow{6}{*}{ImageNet + BCE} &
	    Bone & 0.098 & 0.6622 & 0.3239 & 0.8713 \\
	    & Pleura & 0.4196 & 0.9348 & 0.4478 & 0.9843 \\
	    & Parenchyma & 0.5742 & 0.8351 & 0.6380 & 0.8378 \\
	    & Cardio & 0.4513 & 0.8605 & 0.5617 & 0.9212 \\
	    & Abnormal & 0.6366 & 0.8337 & 0.7323 & 0.7761 \\
	    & {\bf Average} & {\bf 0.4359} & {\bf 0.8253} & {\bf 0.5408} & {\bf 0.887}\\
	\hline
	\multirow {6}{*}{ImageNet + ASL} &
	    Bone & 0.3800 & 0.7123 & 0.2676 & 0.9966 \\
	    & Pleura & 0.4925 & 0.9239 & 0.4925 & 0.9884 \\
	    & Parenchyma & 0.5941 & 0.8389 & 0.5982 & 0.8846 \\
	    & Cardio & 0.5278 & 0.9115 & 0.6255 & 0.9367 \\
	    & Abnormal & 0.6674 & 0.8482 & 0.7123 & 0.8337 \\
	    & {\bf Average} & {\bf 0.5324} & {\bf  0.847} & {\bf 0.5392} & {\bf 0.928}\\
	\hline
	\multirow {6}{*}{ImageNet + DBL} &
	    Bone  & 0.1882 & 0.6993 & 0.1010 & 0.9799 \\
	    & Pleura  & 0.2647 & 0.8691 & 0.403 & 0.9625 \\
	    & Parenchyma & 0.5566 & 0.8195 & 0.6748 & 0.7918 \\
	    & Cardio & 0.3929 & 0.8289 & 0.4723 & 0.9208 \\
	    & Abnormal & 0.6123 & 0.8126 & 0.6993 & 0.7696 \\ 
	    & {\bf Average} & {\bf 0.4029} & {\bf 0.8059} & {\bf 0.4909} & {\bf 0.8894}\\
	\hline
	\multirow {6}{*}{CheXpert + BCE} &
	    Bone  & 0.0706 & 0.5412 & 0.0423 & 0.9962 \\
	    & Pleura  & 0.2623 & 0.8540 & 0.2388 & 0.9867 \\
	    & Parenchyma & 0.537 & 0.7921 & 0.6396 & 0.0.794 \\
	    & Cardio & 0.3872 & 0.8205 & 0.4638 & 0.9208 \\
	    & Abnormal & 0.581 & 0.7789 & 0.6875 & 0.7325 \\
	    & {\bf Average} & {\bf 0.3676} & {\bf 0.7573} & {\bf 0.4144} & {\bf 0.886}\\
	\hline
	\multirow {6}{*}{CheXpert + ASL} &
	    Bone  & 0.4348 & 0.7757 & 0.3521 & 0.9935 \\
	    & Pleura  & 0.5323 & 0.9424 & 0.4925 & 0.9918 \\
	    & Parenchyma & 0.6274 & 0.8624 & 0.6702 & 0.8706 \\
	    & Cardio & 0.5536 & 0.9197 & 0.6043 & 0.9508 \\
	    & Abnormal & 0.6777 & 0.8658 & 0.7512 & 0.8165 \\ 
	    & {\bf {\em Average}} & {\bf {\em 0.5651}} & {\bf {\em 0.8732}} & {\bf {\em 0.5741}} & {\bf {\em 0.9247}}\\
	\hline
	\multirow {6}{*}{CheXpert + DBL} &
	    Bone  & 0.1674 & 0.6912 & 0.2535 & 0.957 \\
	    & Pleura  & 0.4698 & 0.9513 & 0.5224 & 0.984 \\
	    & Parenchyma & 0.5958 & 0.8450 & 0.6104 & 0.8782 \\
	    & Cardio & 0.5094 & 0.9009 & 0.5745 & 0.9422 \\
	    & Abnormal & 0.6498 & 0.8493 & 0.7134 & 0.8100 \\ 
	    & {\bf Average} & {\bf 0.4758} & {\bf 0.8475} & {\bf 0.5349} & {\bf 0.9143}\\
	\hline
\end{tabularx}}
\end{table}

To discover which family of architectures really fits our dataset, we do more experiments with Inception-V3, DenseNet121 and EfficientNet-B2, which are reported to perform well with radiographic images; and two sizes of image 768 and 1024. The result is shown in Table~\ref{table5}, which indicates that bigger image sizes do not give rise to better results, but affect training time. In the matter of model architectures, EfficientNet-B2 outperforms the others. In conclusion, model with EfficientNet-B2 architecture and input size of 768 delivers the best performance.

\begin{table}[!ht]
\centering
\scriptsize{\caption{{\bf Experimental results with different backbones and input sizes.} Model with EfficientNet-B2 architecture and input size of 768 delivers the best performance.}
\label{table5}
	\begin{tabularx}{\linewidth}{|c|*{5}{>{\centering\arraybackslash}X|}}
		\hline
		{\bf Architecture} & {\bf Class} & {\bf F1 score} & {\bf AUC} & {\bf Sensitivity} & {\bf Specificity}\\ 
		{\bf + Input size} & & & & & \\
		\hline
		\multirow{6}{*}{ResNet50 + 768} & 
		    Bone  & 0.4348 & 0.7757 & 0.3521 & 0.9935 \\
		    & Pleura  & 0.5323 & 0.9424 & 0.4925 & 0.9918 \\
		    & Parenchyma & 0.6274 & 0.8624 & 0.6702 & 0.8706 \\
		    & Cardio & 0.5536 & 0.9197 & 0.6043 & 0.9508 \\
		    & Abnormal & 0.6777 & 0.8658 & 0.7512 & 0.8165 \\ 
		    & {\bf Average} & {\bf 0.5651} & {\bf 0.8732} & {\bf 0.5741} & {\bf 0.9247}\\
		\hline
		\multirow{6}{*}{ResNet50 + 1024} & 
		    Bone & 0.3929 & 0.7879 & 0.3099 & 0.9935 \\
		    & Pleura & 0.4593 & 0.9184 & 0.4627 & 0.9874 \\
		    & Parenchyma & 0.6368 & 0.8599 & 0.6549 & 0.8885 \\
		    & Cardio & 0.5521 & 0.9117 & 0.6766 & 0.9342 \\
		    & Abnormal & 0.6856 & 0.8647 & 0.684 & 0.8774 \\
		    & {\bf Average} & {\bf 0.5453} & {\bf 0.8685} & {\bf 0.5576} & {\bf 0.9362}\\
		\hline
		\multirow{6}{*}{DenseNet121 + 768} & 
		    Bone & 0.2087 & 0.7097 & 0.169 & 0.9891 \\
		    & Pleura & 0.4713 & 0.9491 & 0.5522 & 0.9819 \\
		    & Parenchyma & 0.6003 & 0.845 & 0.6656 & 0.8467 \\
		    & Cardio & 0.4965 & 0.8909 & 0.5957 & 0.9317 \\
		    & Abnormal & 0.6524 & 0.8446 & 0.7182 & 0.8096 \\
		    & {\bf Average} & {\bf 0.4858} & {\bf 0.8479} & {\bf 0.5402} & {\bf 0.9118}\\
		\hline
		\multirow{6}{*}{DenseNet121 + 1024} & 
		    Bone & 0.1579 & 0.6892 & 0.1268 & 0.9884 \\
		    & Pleura & 0.3515 & 0.9292 & 0.6269 & 0.9557 \\
		    & Parenchyma & 0.5777 & 0.8298 & 0.6212 & 0.8531 \\
		    & Cardio & 0.4624 & 0.872 & 0.5362 & 0.9335 \\
		    & Abnormal & 0.6334 & 0.8323 & 0.7252 & 0.7775 \\
		    & {\bf Average} & {\bf 0.4366} & {\bf 0.8305} & {\bf 0.5272
} & {\bf 0.9016}\\
		\hline
		\multirow{6}{*}{InceptionV3 + 768} & 
		    Bone & 0.2529 & 0.7198 & 0.1549 & 0.9983 \\
		    & Pleura & 0.45 & 0.9392 & 0.5373 & 0.9806 \\
		    & Parenchyma & 0.6015 & 0.8429 & 0.6227 & 0.8757 \\
		    & Cardio & 0.556 & 0.901 & 0.5702 & 0.9591 \\
		    & Abnormal & 0.6606 & 0.8476 & 0.6899 & 0.843 \\
		    & {\bf Average} & {\bf 0.5042} & {\bf 0.8501} & {\bf 0.515} & {\bf 0.9313}\\
		\hline
		\multirow{6}{*}{InceptionV3 + 1024} & 
		    Bone & 0.3364 & 0.7454 & 0.2535 & 0.9939 \\
		    & Pleura & 0.4379 & 0.9282 & 0.5522 & 0.9778 \\
		    & Parenchyma & 0.6037 & 0.8386 & 0.6319 & 0.8719 \\
		    & Cardio & 0.527 & 0.8861 & 0.4979 & 0.9667 \\
		    & Abnormal & 0.6447 & 0.8411 & 0.658 & 0.849 \\
		    & {\bf Average} & {\bf 0.5099} & {\bf 0.8479} & {\bf 0.5187} & {\bf 0.9319}\\
		\hline
		\multirow{6}{*}{EfficientNetB2 + 768} & 
		    Bone & 0.4483 & 0.8035 & 0.3662 & 0.9935 \\
		    & Pleura & 0.5085 & 0.9567 & 0.4478 & 0.9928 \\
		    & Parenchyma & 0.6354 & 0.8602 & 0.6764 & 0.8744 \\
		    & Cardio & 0.5597 & 0.9196 & 0.6085 & 0.9519 \\
		    & Abnormal & 0.6970 & 0.8671 & 0.6946 & 0.8825 \\
		    & {\bf {\em Average}} & {\bf {\em 0.5698}} & {\bf {\em 0.8814}} & {\bf {\em 0.5587}} & {\bf {\em 0.9390}}\\
		\hline
		\multirow{6}{*}{EfficientNetB2 + 1024} & 
		    Bone & 0.3704 & 0.7600 & 0.3521 & 0.9867 \\
		    & Pleura & 0.5075 & 0.9352 & 0.5075 & 0.9888 \\
		    & Parenchyma & 0.6329 & 0.8635 & 0.7178 & 0.8472 \\
		    & Cardio & 0.5132 & 0.9129 & 0.6596 & 0.9226 \\
		    & Abnormal & 0.6793 & 0.8595 & 0.6745 & 0.8774  \\
		    & {\bf Average} & {\bf 0.5407} & {\bf 0.8662} & {\bf 0.5823} & {\bf 0.9245}\\
		\hline
	\end{tabularx}}
\end{table}

Detailed result of our best model is also presented in Table \ref{table6}. By using ASL, the chest wall class has improved significantly when increasing to nearly 32\% compared to the model using BCE and not using CheXpert as pre-trained. The pleura class has less samples than the chest wall, but the results do not improve much after using ASL, possibly because the chest wall class has a more diverse number of abnormal manifestations in our data, so the model focused more on this class.

\begin{table}[!ht]
\centering
\scriptsize{\caption{{\bf Performance of EfficientNet-B2 on five classes.}}
\label{table6}
\begin{tabularx}{\linewidth}{|c|*{4}{>{\centering\arraybackslash}X|>{\centering\arraybackslash}X|}}

\hline
{\bf Class} & {\bf Macro F1 score (95\%CI)} & {\bf Average AUC (95\%CI)} & {\bf Average Sensitivity} & {\bf Average Specificity} \\ 
\hline

\multirow{2}{*}{Chest wall} & 0.4483 & 0.8035 & 0.3662 & 0.9935 \\ 
& (0.327, 0.559) & (0.748, 0.857) & & \\
\hline

\multirow{2}{*}{Pleura} & 0.5085 & 0.9567 & 0.4478 & 0.9928 \\ 
& (0.387, 0.617) & (0.938, 0.973) & & \\
\hline

\multirow{2}{*}{Parenchyma} & 0.6354 & 0.8602 & 0.6764 & 0.8744 \\ 
& (0.606, 0.664) & (0.843, 0.876) & & \\
\hline

\multirow{2}{*}{Cardio} & 0.5597 & 0.9196  & 0.6085 & 0.9519 \\ 
& (0.507, 0.608) & (0.902, 0.936) & & \\
\hline

\multirow{2}{*}{Abnormal} & 0.6970 & 0.8671 & 0.6946 & 0.8825 \\ 
& (0.673, 0.722) & (0.853, 0.882) & & \\
\hline

\end{tabularx}}
\end{table}

Fig~\ref{fig6} illustrates plots on all tasks. The model achieves the best AUC on pleura class (0.96), and the worst on chest wall class (0.81). The abnormal class recorded 0.87 AUC, the parenchyma and cardiac classes witness figures of 0.86 and 0.92, respectively.

\begin{figure}[!ht]
\centering
\includegraphics[width=0.48\linewidth, scale=0.2]{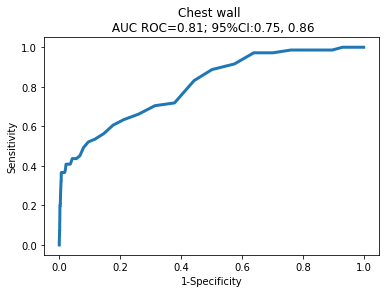}
\includegraphics[width=0.48\linewidth, scale=0.2]{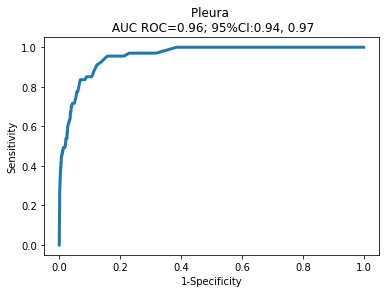}
\includegraphics[width=0.48\linewidth, scale=0.2]{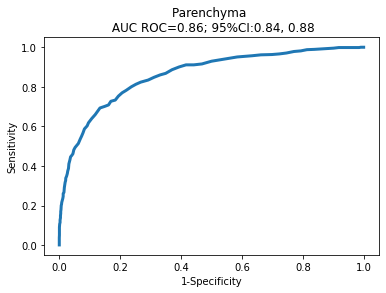}
\includegraphics[width=0.48\linewidth, scale=0.2]{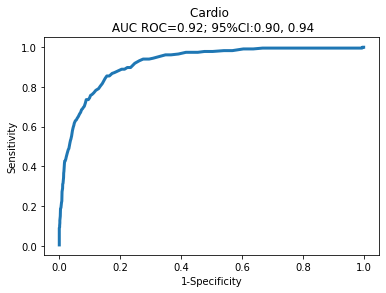}
\includegraphics[width=0.48\linewidth, scale=0.2]{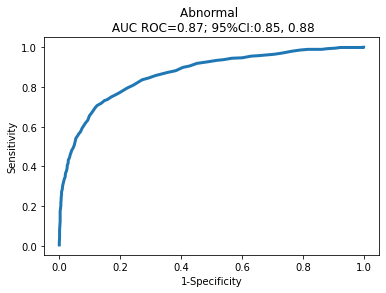}
\caption{{\bf Area under the ROC Curve.} Pleura class delivered the highest AUC value, at 0.96 (95\% CI 0.94, 0.97) whereas chest wall class performed the lowest AUC value, with the figure of 0.81 (95\% CI 0.75, 0.85).}
\label{fig6}
\end{figure}

The same procedure is also applied to build the two models of fine classification (detection of 14 pathologies) and coarse classification (detection of abnormalities in 4 locations in CXR images), in order to evaluate the effectiveness of the coarse classification compared to the fine classification. We use the CheXpert benchmark dataset to build and evaluate two models sharing the same configurations to retain the sense of objectivity. The data in the CheXpert dataset are labeled with 14 classes corresponding to 13 abnormalities in the chest radiograph and an implication of no findings. We infer where the lesion is in the 4 considered positions based on the type of lesion indicated in the CheXpert dataset. Table~\ref{tab:lookup} shows the mappings between CheXpert data labels (14 classes) and the proposed set of labels (5 classes). Comparison of coarse and fine classification on Table~\ref{table7}. Based on the results shown in the Table~\ref{table7}, it can be seen that the coarse classification method gives a higher F1 score in both the abnormal class and the macro average F1 score.

\begin{table}[!ht]
	\centering
	\scriptsize{
	\caption{\textbf{The mappings between CheXpert data labels (14 classes) and the proposed set of labels (5 classes).} P and N refer to positive and negative respectively}
	\label{tab:lookup}
    \begin{tabular}{|c|c|c|c|c|c|}
    \hline 
      & \textbf{Chest wall} & \textbf{Pleura} & \textbf{Parenchyma} & \textbf{Cardio} & \textbf{Abnormal} \\
    \hline 
     No Finding & N & N & N & N & N \\
    \hline 
     Enlarged Cardiom & N & N & N & \textbf{P} & \textbf{P} \\
    \hline 
     Cardiomegaly & N & N & N & \textbf{P} & \textbf{P} \\
    \hline 
     Lung Lesion & N & N & \textbf{P} & N & \textbf{P} \\
    \hline
     Lung Opacity & N & N & \textbf{P} & N & \textbf{P} \\
    \hline 
     Edema & N & N & \textbf{P} & N & \textbf{P} \\
    \hline 
     Consolidation & N & N & \textbf{P} & N & \textbf{P} \\
    \hline 
     Pneumonia & N & N & \textbf{P} & N & \textbf{P} \\
    \hline 
     Atelectasis & N & N & \textbf{P} & N & \textbf{P} \\
    \hline 
     Pneumothorax & N & \textbf{P} & N & N & \textbf{P} \\
    \hline 
     Pleural Effusion & N & \textbf{P} & N & N & \textbf{P} \\
    \hline 
     Pleural Other & N & \textbf{P} & N & N & \textbf{P} \\
    \hline 
     Fracture & \textbf{P} & N & N & N & \textbf{P} \\
    \hline 
     Support Devices & N & N & N & N & \textbf{P} \\
     \hline
    \end{tabular}}
\end{table}

\begin{table}[!ht]
    \centering
    \scriptsize{\caption{{\bf Comparison of coarse and fine classification on CheXpert.}}
    \label{table7}
    \begin{tabularx}{\linewidth}{|>{\centering\arraybackslash}X|*{4}{>{\centering\arraybackslash}X|}}
         \hline \multirow{2}{*}{{\bf Architecture}} & \multicolumn{2}{c|}{{\bf 5 classes}} & \multicolumn{2}{c|}{{\bf 12 classes}} \\
         \cline{2-5} & {\bf Macro F1 score} & {\bf F1 score on Abnormal class} & {\bf Macro F1 score} & {\bf F1 score on Abnormal class} \\
         \hline ResNet50~\cite{bib32}      & 0.7109  & 0.9443           & 0.4849   & 0.9444 \\
         \hline DenseNet121~\cite{bib33}   & 0.7208  & 0.9519           & 0.4650   & 0.9438 \\
         \hline InceptionV3~\cite{bib34}   & 0.7181  & 0.9491           & 0.4846   & 0.9492 \\
         \hline {\bf EfficientB2~\cite{bib35}}   & {\bf 0.7429} & {\bf 0.9520}           & {\bf 0.5044}   & {\bf 0.9450} \\
         \hline
    \end{tabularx}}
\end{table}

We also plot Grad-CAMs~\cite{bib28} to give the visual explanations of how the model fulfil predictions. Fig~\ref{fig7} illustrates the original images and their respective Grad-CAMs. In both cases, the pathologies in the collarbone (nondisplaced fracture) and in the pleura (pleural effusion) were correctly highlighted. The results are attained when performing with the EfficientNet-B2 architecture, the input size is 768x768, using the CheXpert dataset to build the pretrained model and apply the asymmetric loss function.

\begin{figure}[!ht]
\centering
\includegraphics[width=0.24\linewidth, scale=0.2]{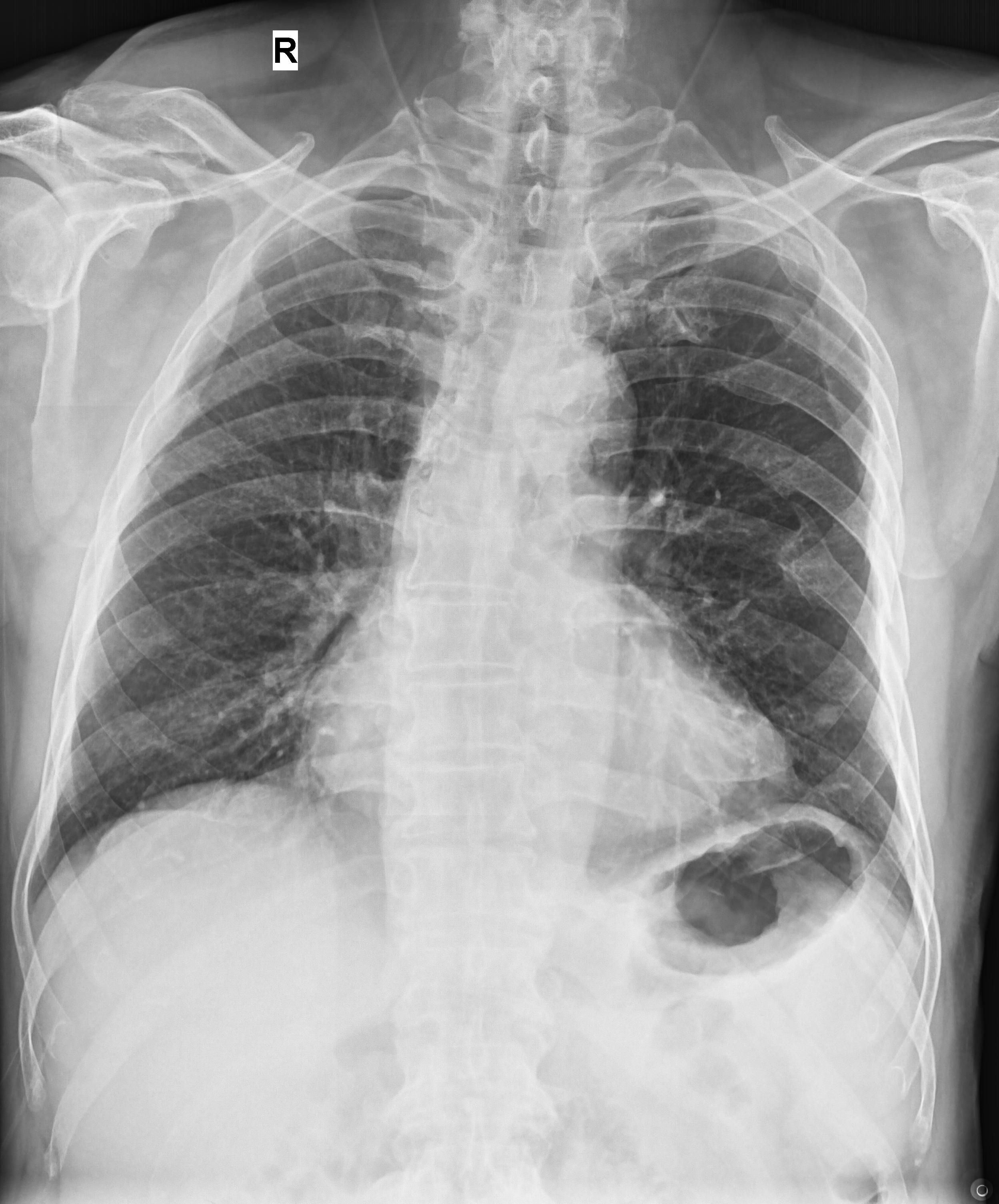}
\includegraphics[width=0.24\linewidth, scale=0.2]{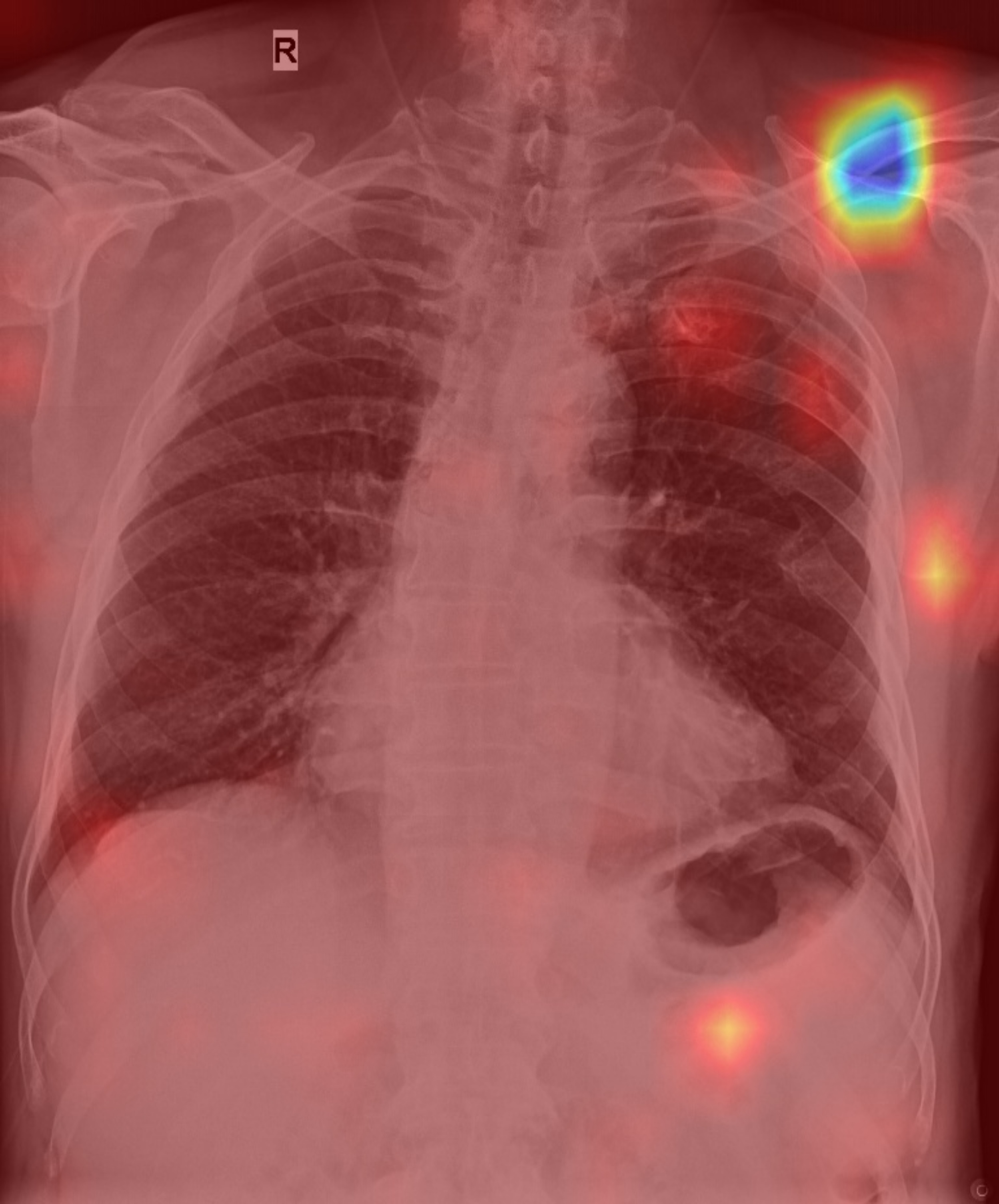}
\includegraphics[width=0.24\linewidth, scale=0.2]{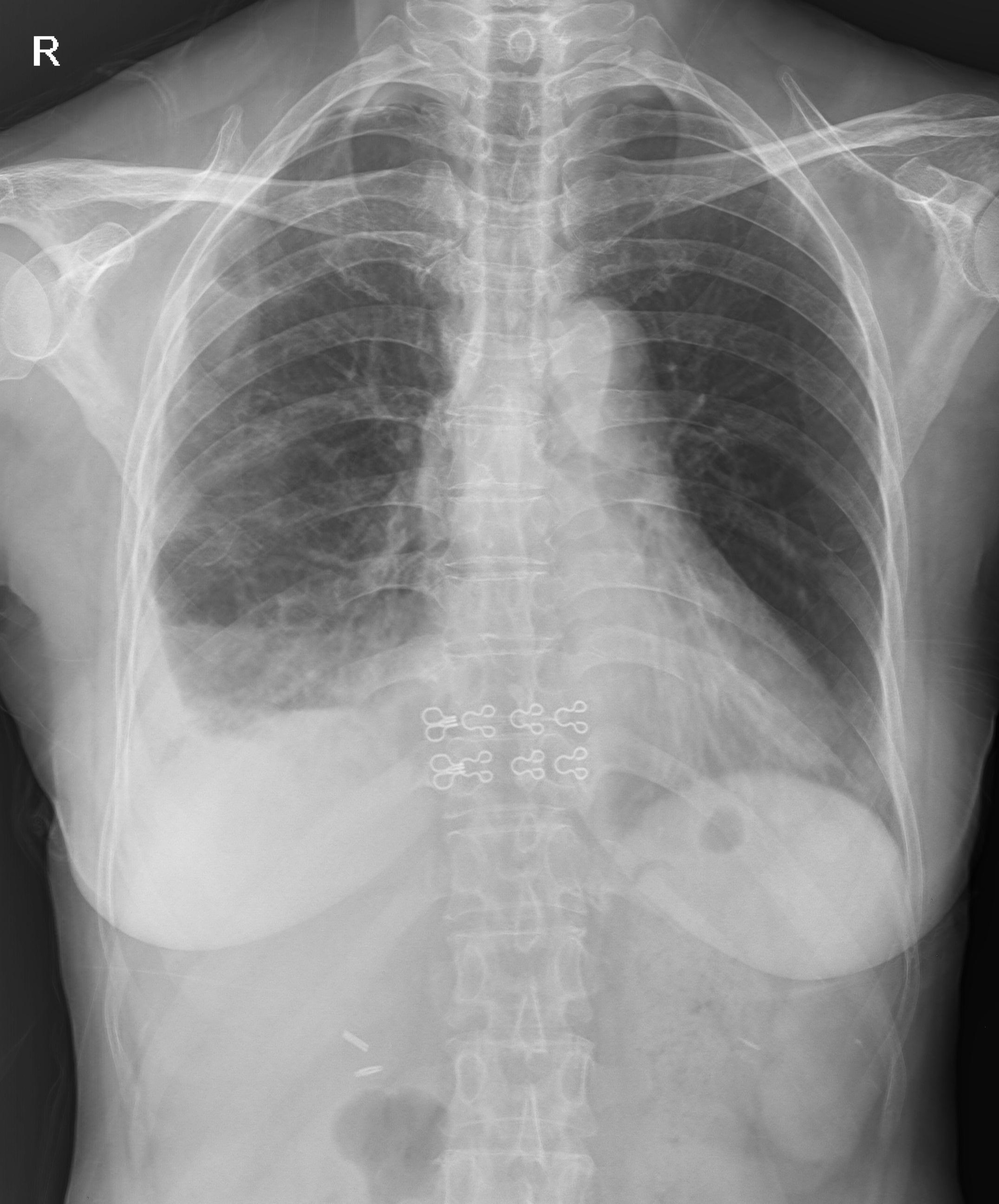}
\includegraphics[width=0.24\linewidth, scale=0.2]{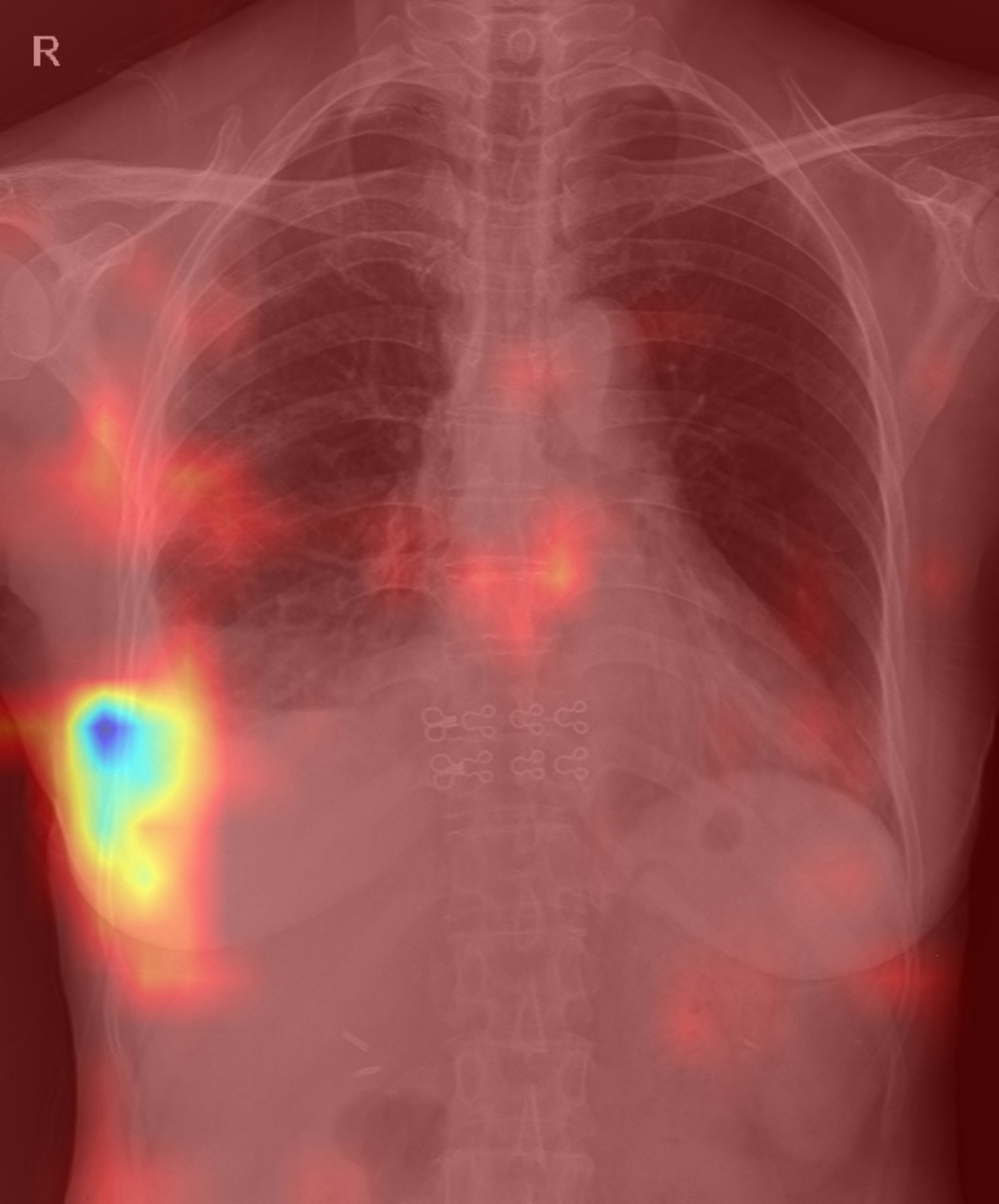}
\caption{{\bf Original images and respective Grad-CAMs.} There is a collarbone (nondisplaced fracture) in the first two figures, while the last two ones containing pleural effusion in the pleura. Both of these pathologies were correctly highlighted.}
\label{fig7}
\end{figure}

\newpage

\section*{Conclusion}

In current work, we propose a semi-automatic process of building an accurate CXR dataset, which can take advantage of the resources stored in PACS and HIS systems, especially minimizing the intervention of radiologists. We also suggest a coarse classification method based on the location of abnormalities in radiographs, which can address the realistic demand for Vietnamese radiologists and be more efficient than classification based on pathology types. Finally, we demonstrate that building pre-trained models using large CXR datasets can significantly improve performance compared to using ImageNet datasets. The models finetuned  from CheXpert pre-trained models with asymmetric loss function achieve significant gains over ImageNet pre-trained models, which we believe will serve as a strong baseline for future research. We also believe that this method will be applied for other languages which have the same characteristic and task requirement.




\newpage


\end{otherlanguage}
\end{document}